\documentclass[superscriptaddress,secnumarabic,amssymb,amsmath,nobibnotes,aps,prd,showkeys,nofootinbib,preprint]{revtex4}

\usepackage{graphicx}
\usepackage{float}
\usepackage{bm}
\usepackage{amsmath}
\usepackage{amsfonts}
\usepackage{amssymb}
\usepackage{epstopdf}
\usepackage{natbib}%
\newcommand{\bee}{\begin{equation}}
\newcommand{\eee}{\end{equation}}
\newcommand{\eaa}{\end{eqnarray}}
\newcommand{\baa}{\begin{eqnarray}}
\def\ni{\noindent}
\def\noo{\nonumber \\}

\usepackage{color}

\begin{document}
	

\title{On the nature of R\'enyi modified entropy and the Incomplete statistics approach in black holes thermodynamics}

\author{Everton M. C. Abreu}\email{evertonabreu@ufrrj.br}
\affiliation{Departamento de F\'{i}sica, Universidade Federal Rural do Rio de Janeiro, 23890-971, Serop\'edica, RJ, Brazil}
\affiliation{Departamento de F\'{i}sica, Universidade Federal de Juiz de Fora, 36036-330, Juiz de Fora, MG, Brazil}
\affiliation{Programa de P\'os-Gradua\c{c}\~ao Interdisciplinar em F\'isica Aplicada, Instituto de F\'{i}sica, Universidade Federal do Rio de Janeiro, 21941-972, Rio de Janeiro, RJ, Brazil}
\author{Jorge Ananias Neto}\email{jorge@fisica.ufjf.br}
\affiliation{Departamento de F\'{i}sica, Universidade Federal de Juiz de Fora, 36036-330, Juiz de Fora, MG, Brazil}


\begin{abstract}
\noindent {In this work we have investigated the effects of the two highlighted nongaussian entropies which are the modified R\'enyi entropy and the so-called Incomplete statistics in the analysis of the thermodynamics of black holes (BHs). We have obtained the equipartition theorems and after that we obtained the heat capacities for both approaches. Depending on the values of  both $\lambda-$parameter and $M$, the BH mass, relative to a modified R\'enyi entropy and, depending on the values of the $q-$parameter and $M$, the Incomplete entropy can determine if the BH has an unstable thermal equilibrium or not for each model.}

\end{abstract}

\maketitle

{\color{black} 
\section{Introduction}

Hawking's \cite{swh} discovery of a thermal radiation from a black hole (BH) was unexpected to the majority of specialists at that time.  This surprise occurred even though the existence of quite a few demonstrations that a closed connection between BH physics and thermodynamics had appeared before Hawking's paper. Two years before that paper, Bekenstein \cite{jdb} has realized that the properties of one of the BH parameters, its area, for example, resemble the concept of entropy.  In fact, Hawking's area theorem \cite{swh} means that the area $A$ does not decrease in any classical scenario.   Namely, it behaves just like the entropy does. It was found, as a matter of fact, that the similarity between BH physics and thermodynamics is quite general. It deals with {\it gedanken} mental exercises where we have specific thermodynamic devices and with the general laws of thermodynamics.  Both of them have analogous ideas in BH physics.  An arbitrary BH, like a well established thermodynamic system, enters in an equilibrium (stationary) state after the relaxation processes are finished.

{ Among some extensions of Boltzmann-Gibbs (BG) statistical mechanics that were proposed during the last decades, we can mention the Tsallis entropy \cite{tsallis}, which is defined as
\begin{eqnarray}
	\label{tsad}
	S_T = k_B \,\frac{1}{1-q} \, \sum_{i=1}^W \left(p_i^q - p_i\right) \,\,,
\end{eqnarray}

\ni where $W$ is the number of discrete configurations,  $p_i$ denotes the ordinary probability of accessing state $i$ and $q$ is a real parameter that measures the degree of nonextensivity. The definition of entropy in Tsallis statistics carries the standard properties of positivity, equiprobability, concavity and irreversibility. This approach has been successfully used in many different physical systems. For instance, we can mention the Levy-type anomalous diffusion \cite{levy}, turbulence in a pure-electron plasma \cite{turb} and gravitational systems \cite{sys,sa,eu,maji,mora2}.
It is noteworthy to mention that Tsallis thermostatistics formalism has the BG statistics as a particular case in the limit $ q \rightarrow 1$ where the standard additivity of entropy can be recovered.

In addition, there is another important $q$-generalized entropy defined as
\begin{eqnarray}
	\label{renyi}
	S_R = k_B \,\frac{1}{1-q} \, \ln \sum_{i=1}^W \, p_i^q \,\,,
\end{eqnarray}
}
{
\ni which is known as R\'enyi entropy \cite{renyi}. Combining both Eqs. (\ref{renyi}) and (\ref{tsad}),} { and using the usual normalization
condition $\sum^W_{i=1} p_i=1$ we have
\begin{eqnarray}
	\label{ret}
	S_R= k_B \,\frac{1}{\lambda} \, \ln (1+\lambda \frac{S_T}{k_B}) \,\,,
\end{eqnarray}
}
{
\ni where  $\lambda$ ($\lambda \equiv 1-q$) is also a constant parameter. If we take the limit $\lambda\rightarrow 0$ in Eq.  (\ref{ret}) then we have 
$S_R=S_T$ where $S_T$ is the Tsallis entropy as defined in Eq. (\ref{tsad}).
This modified R\'enyi entropy (MRE) model was suggested initially  by Bir\'o and Czinner \cite{ci}.  It was also used by other authors for example in ref. \cite{ko,many}. Following Bir\'o and Czinner, that considered Tsallis entropy, $S_T$, as the BH entropy, $S_{BH}$, and they wrote the MRE as a function of $S_{BH}$. The final form is given by
\begin{eqnarray}
	\label{sar}
	S_R= k_B \,\frac{1}{\lambda} \, \ln (1+\lambda \frac{S_{BH}}{k_B}) \,\,.
\end{eqnarray}
}

Having said that, the so-called Incomplete statistics (IS) \cite{inco1,lbs}, analogously to Tsallis thermostatistics formalism  and { R\'enyi entropy}, generalizes the usual BG statistics since this last one is not adequate to deal with more complicated physical phenomena. Among them we can mention for example fractal and self-similar frameworks, long-range interactions, long-duration memory, anomalous diffusion phenomena and Loop Quantum Gravity. The IS entropy can be directly used to obtain a new equation for the probability distribution. The  objective is to overcome problems in BG probability distribution.
These statistical formalisms constitute the main part of the so-called  nonextensive statistical mechanics, and we know that the most difficult physical systems are frequently or usually nonextensive. 
Some applications related to IS can be found in references \cite{todos,nos1}. A normalization  condition adopted in IS is
\begin{eqnarray}
\label{inn}
\sum_{i=1}^W p_i^q = 1 \,\,,
\end{eqnarray}

\ni where $p_i$ is the probability of the system to be in a $i$-microstate, $W$ is the total number of configurations and $q$ is known as the $q$-parameter, that measures the nonextensivity of the system. At the limit $q \rightarrow 1$ we must  recover the standard normalization condition which is given by $\sum_{i=1}^W p_i = 1$. 
Adopting the Wang condition, which is given in Eq.  (\ref{inn}), the entropy in Eq.  (\ref{tsad}) can be written as
\begin{eqnarray}
\label{inss}
S_q=k_B \frac{1 - \sum_{i=1}^W p_i}{1-q} \,\,,
\end{eqnarray}

\ni where $q>0$ is required by the incomplete normalization in Eq.  (\ref{inn}).
Using the microcanonical ensemble definition, where all the states have the same probability and consequently, due to the normalization condition in Eq.  (\ref{inn}), we have that $p_i^q=1/W$ and the IS entropy \cite{inco1} reduces to 
\begin{eqnarray}
\label{inmicro}
S_q=k_B \, \frac{W ^{\frac{(q-1)}{q}}-1}{q-1} \,\,,
\end{eqnarray}

\ni where at the limit $q \rightarrow 1$,  we must recover the usual BG entropy formula, i.e., $S=k_B\, \ln {W}$.

\section{Black hole thermodynamics}

It is well known that the thermodynamics of BHs is based on the concepts of both entropy and temperature of a BH \cite{jdb,swh}. The temperature of a BH horizon is directly proportional to its surface gravity. In Einstein gravitation theory, the horizon entropy of a BH is proportional to its horizon area, i.e., the entropy area law of a BH. From now on we will use that $\hbar=c=k_B=1$.

Our first issue \cite{ci,nos2} will be the Schwarzschild BH entropy which is written as 
\begin{eqnarray}
\label{sm-0}
S_{BH}= 4 \pi G M^2 \,,
\end{eqnarray}

\ni where $G$ is the gravitational constant and $M$ is the mass of BH. The temperature is given by
\begin{eqnarray}
\label{sm}
\frac{1}{T}=\frac{\partial S(M)}{\partial M} \,,
\end{eqnarray}

\ni and using Eq.  \eqref{sm-0} we have that
\begin{eqnarray}
\label{ts2}
\frac{1}{T}=\frac{\partial S_{BH}(M)}{\partial M}=8 \pi G M \,.
\end{eqnarray}

\ni Moreover, the number $N$ of degrees of freedom (DF) in the horizon can be given by assuming the relation \cite{ko}
\begin{eqnarray}
\label{ns}
N=4S\,,
\end{eqnarray}

\ni where $S$ is an specific entropy describing the horizon. So, using Eq.  (\ref{ns}) in our initial case, we have 
\begin{eqnarray}
\label{ns2}
N=16\pi G M^2 \,.
\end{eqnarray}

{\color{black} 
\ni Combining Eqs. (\ref{ts2}) and (\ref{ns2}) and making some algebra then we can derive the usual equipartition theorem \cite{nos2,nos3,nos4,nos5}
\begin{eqnarray}	
\label{eqs1}
M=\frac{1}{2} N T \,,
\end{eqnarray}
}
\ni which corresponds to the horizon energy.

To establish the physical coherence, a standard test is to calculate the heat capacity of the model.   The sign of the heat capacity can support us in determining the stable thermal equilibrium of a BH.  Namely, a positive heat capacity is meaningful.   On the other hand, a negative heat capacity in such system shows a thermodynamical unstableness.

The heat capacity can be computed from the expression
\bee
\label{heat-capacity}
C\,=\,-\,\frac{[S_{BH}^{\;\prime}(M)]^2}{S_{BH}^{\;''}(M)} \,\,,
\eee

\ni where each prime means a single derivative relative to $M$. So, substituting the entropy of Eq.  \eqref{sm-0} into Eq.  \eqref{heat-capacity} we have that 

\bee
\label{h-capacity}
C_{BH}\,=\,-\,8 \pi G M^2 \,\,,
\eee

\ni which means that due to the negative value of Eq.  (\ref{h-capacity}) the Schwarzschild BH { has an unstable thermal equilibrium.}

{
\section{The modified R\'enyi entropy and black holes thermodynamics}

The first model that we analyzed is the MRE. Using Eq.  (\ref{sm-0}) into Eq.  (\ref{sar}) we have
\begin{eqnarray}
	\label{sarm}
	S_R= \frac{1}{\lambda} \, \ln \Big( 1+4 \pi \lambda G M^2 \Big) \,\,.
\end{eqnarray}

\ni Using Eqs. (\ref{sarm}) and (\ref{ns}) we obtain, respectively, the black hole temperature and the number $N$ of DF as
\begin{eqnarray}
	\label{tsr}
	T =\frac{1+4 \pi \lambda G M^2}{8 \pi G M} \,\,,
\end{eqnarray}

\ni and
\begin{eqnarray}
	\label{nr}
	N=\frac{4}{\lambda} \ln \Big(1+4 \pi \lambda G M^2 \Big) \,.
\end{eqnarray}

\ni Using Eqs. (\ref{tsr}) and (\ref{nr}), and after some algebra, we have the equation
\begin{eqnarray}
	\label{eqsr}
	M^2\bigg(1+\frac{1}{e^{\frac{N\lambda}{4}} - 1}\bigg)-\frac{2T}{\lambda}\,M =0 \,\,,
\end{eqnarray}

\ni  and its non-zero solution is the mathematical expression for the equipartition theorem which is compatible with the MRE, namely, 
\begin{eqnarray}
	\label{eqsr2}
	M=\frac{2 T}{\lambda}  \frac{e^{\frac{N\lambda}{4}} - 1}{e^{\frac{N\lambda}{4}}}\,\,.
\end{eqnarray}

\ni It is straightforward to show that if we take the limit $\lambda\rightarrow 0$ in  Eq.  (\ref{eqsr2}) then we recover the usual equipartition law, 
$M= 1/2 N T$. 

The heat capacity, using Eqs. \eqref{sarm} and \eqref{heat-capacity} is 
\bee
\label{heat-cap-mod-renyi}
C_R \,=\,\frac{8\pi G M^2}{4\pi \lambda G M^2 -1} \,\,,
\eee

\ni and when we make $\lambda=0$ in Eq. (\ref{heat-cap-mod-renyi}) we recover the heat capacity written in Eq.  (\ref{h-capacity}). Here, it is important to mention that 
Eqs. (\ref{tsr}) and (\ref{heat-cap-mod-renyi}) have already been obtained in ref. \cite{ci}.

From Eq.  (\ref{tsr}) we can derive a temperature variational condition \cite{ks} such that
\begin{eqnarray}
	\label{vctr}
	\frac{\partial T}{\partial M}\Bigg|_{M=M_{ext}} = 0 \; \qquad \Rightarrow \;\qquad 4 \pi G M^2_{ext} = \frac{1}{\lambda}\,\,,
\end{eqnarray}

\ni and Eq.  (\ref{vctr}) determines that the allowed values for the $\lambda$-parameter, since the term $8 \pi G M^2_{ext}$ is always positive, is $\lambda > 0$.

So, substituting Eq.  (\ref{vctr}) into (\ref{tsr}) we can write the Hawking temperature as
\begin{eqnarray}
	\label{tvc}
	T =  \frac{\lambda}{2}\,  \left( \frac{M_{ext}^2}{M} +M \right) \,\,,
\end{eqnarray}

\ni  and making use of Eq.  (\ref{vctr}), and substituting  $\lambda=0$ into Eq. (\ref{tvc}), then we recover the usual Hawking temperature, Eq.  (\ref{ts2}).
In Fig. 1, the temperature in Eq.  \eqref{tvc} was plotted as a function of mass $M$ for $\lambda=1/2$ and from Eq.  (\ref{vctr})
and making $G=1$ we have $M_{ext}\cong 0.4$.
\begin{figure}[H]
	\centering
	\includegraphics[width=8.cm]{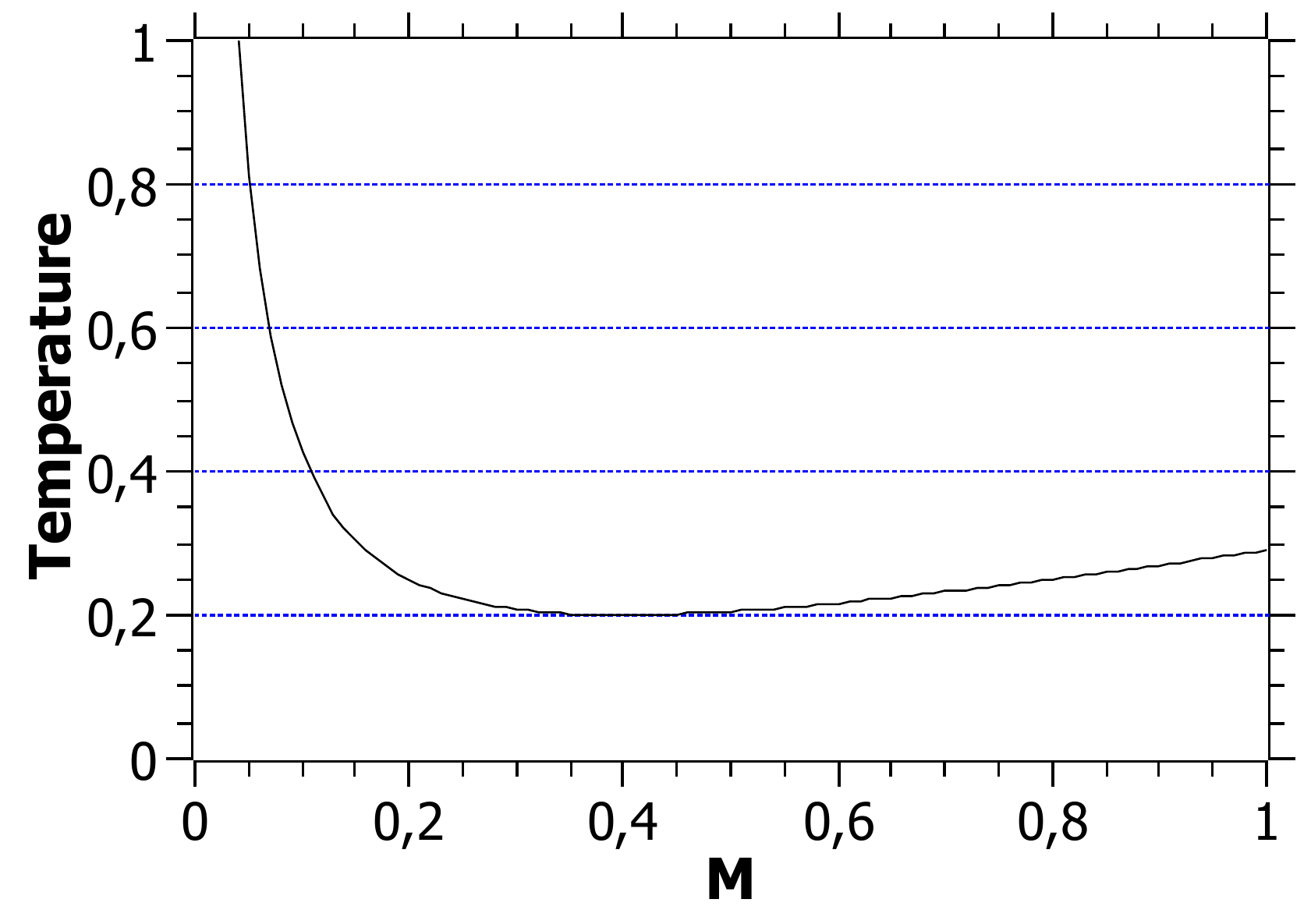}
	\caption{BH temperature for modified R\'enyi entropy, Eq. (\ref{tvc}), as a function of mass $M$ for $\lambda=1/2$, $G=1$
		and $M_{ext}=0.4$.}
\end{figure}

\ni From Fig. 1 we can observe that for $M=0.4$ we have the minimum value of the temperature.
The heat capacity, using Eqs. (\ref{sarm}) and (\ref{heat-capacity}), is
\begin{eqnarray}
	\label{cre}
	C_R=- \, \frac{8 \pi G M^2}{4 \pi \lambda G M^2-1}       \,.
\end{eqnarray}

\ni  When we make $\lambda=0$ in Eq. (\ref{cis}) we recover the usual value of the Schwarzschild BH heat capacity which is Eq.  (\ref{h-capacity}).
 So, substituting Eq.  (\ref{vctr}) into Eq. (\ref{cre}) we can write the heat capacity as
\begin{eqnarray}
	\label{crev}
	C_R^{ext}= \, \frac{2}{\lambda} \, \frac{\frac{M^2}{M^2_{ext}}}{\frac{M^2}{M^2_{ext}}-1}       \,.
\end{eqnarray}

\ni From Eq.  (\ref{crev}) we can observe that for $ M < M_{ext}$ the heat capacity of system is negative. Consequently the BH has an unstable thermal equilibrium.
For $ M > M_{ext}$ the heat capacity of system is positive and the BH has a stable thermal equilibrium. In Fig. 2, the heat capacity, Eq.  (\ref{crev}), was plotted as a function of mass $M$ for $\lambda=1/2$ and $M_{ext}=0.4$. So, we can observe that for $M=M_{ext}=0.4$ the heat capacity diverges. This result can indicate a possible phase transition between a thermally unstable phase and a thermally stable phase of the BH considering the modified R\'enyi entropy.

\begin{figure}[H]
	\centering
	\includegraphics[width=8.cm]{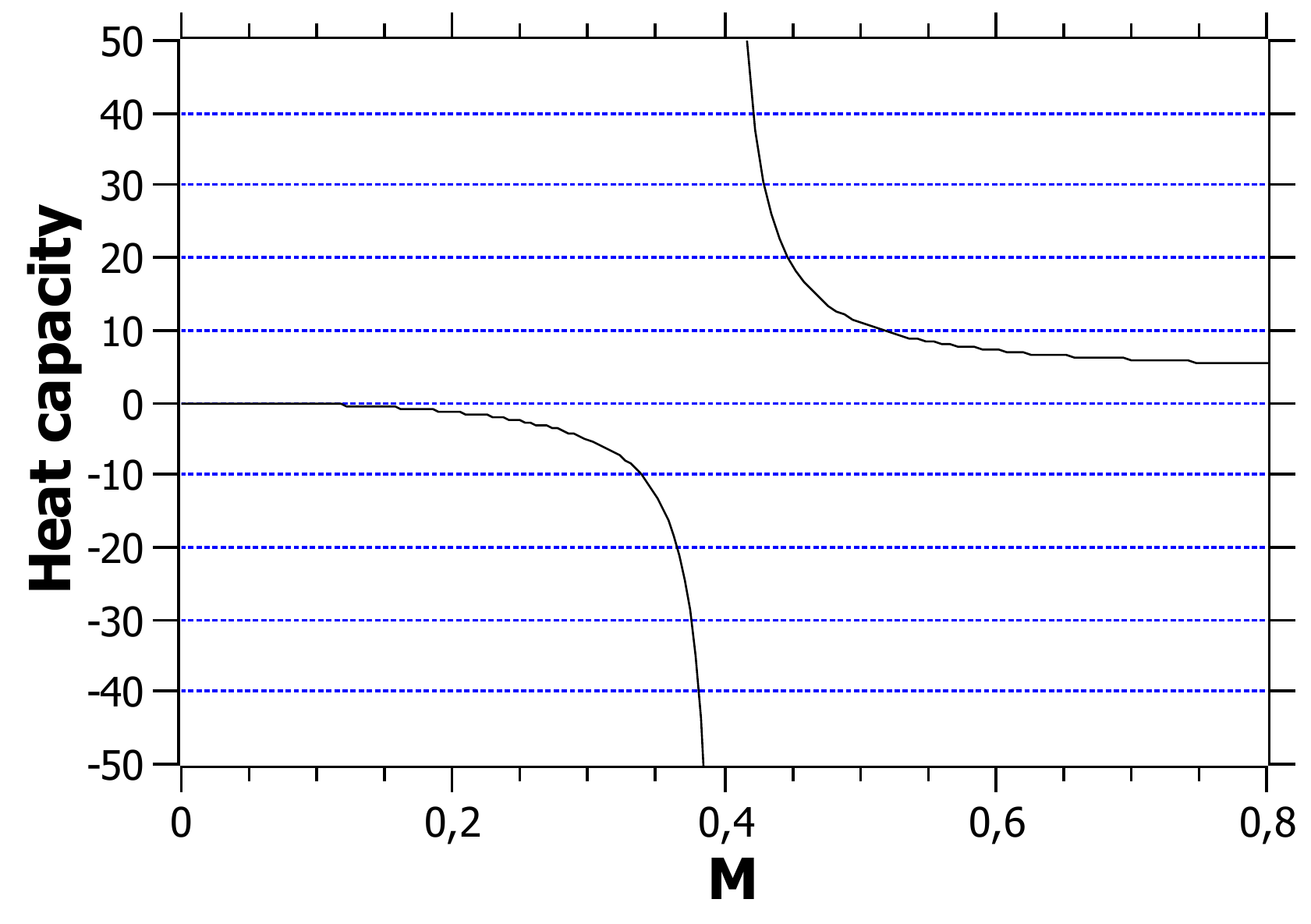}
	\caption{BH heat capacity of modified R\'enyi entropy, Eq. (\ref{crev}), as a function of mass $M$. We have used $\lambda=1/2$ 
		and $M_{ext}=0.4$.}
\end{figure}
}

\section{The Incomplete statistics and black holes thermodynamics}

{ R\'enyi and Tsallis statistics are two important approaches that show very positive results. More specifically, these two formalisms have led to
significant contributions in the analysis of BH thermodynamics \cite{ci,ks}. Since IS statistics has the same theoretical basis of Tsallis' statistics, except
in the normalization condition, Eq.  (\ref{inn}), then it would also be interesting to apply the IS entropy in the investigation of BH thermodynamics.}

Our procedure begins by considering that the BG entropy, $S= \ln W$, describes the BH entropy, 
Eq.  (\ref{sm-0}) \cite{ks,mora}
\bee
\label{bg}
\ln W = 4 \pi G M^2 \qquad \Longrightarrow \qquad W = \exp \Big( 4 \pi G M^2\Big) \,.
\eee

\ni Now, using Eq.  (\ref{bg}) into Eq. (\ref{inmicro}) we find
\begin{eqnarray}
\label{isbh}
S_q= \frac{1}{q-1}\,\Bigg[\exp\Bigg(\frac{4 \pi (q-1) G M^2}{q} \,\Bigg)-1\Bigg] \,.
\end{eqnarray}

\ni Making use of Eqs. (\ref{sm}) and (\ref{ns}) we can obtain, respectively, the Hawking temperature and the
number $N$ of DF as 
\begin{eqnarray}
\label{tbh}
\frac{1}{T} = \frac{8\pi G M}{q} \, \exp \Bigg(\frac{4 \pi (q-1) G M^2}{q} \,\Bigg) \,,
\end{eqnarray}

\ni and 
\begin{eqnarray}
\label{nis}
N = \frac{4}{q-1} \, \,\Bigg[\exp\Bigg(\frac{4 \pi (q-1) G M^2}{q} \,\Bigg)-1\Bigg]     \,.
\end{eqnarray}

\ni Combining Eqs. (\ref{tbh}) and (\ref{nis}) and after some algebra then we can obtain an equipartition theorem for the BH mass in the IS as
\begin{eqnarray}
\label{eqis}
M= \frac{q}{2\pi G} \, \frac{1}{\left[ 4+ (q-1) N \right] T}  \,.
\end{eqnarray}

\ni When we make $q=1$ in Eq. (\ref{eqis}) we recover the usual relation for both temperature and mass in Schwarzschild BH, which is Eq.  (\ref{ts2}).   An alternative way would be from Eq.  \eqref{tbh} where we have that
\bee
\label{A}
q\,=\,8\pi GMT e^{\delta M^2} \,,
\eee

\ni where, conveniently and momentarily,  $\delta=4\pi G(q-1)/q$.  Notice that $q \rightarrow 1$ is equivalent to $\delta \rightarrow 0$.   Let us show that we have the standard equipartition law in the $q \rightarrow 1$ limit.  

From Eq.  \eqref{nis} we can write that
\bee
\label{B}
q-1\,=\,\frac 4N \Big( e^{\delta M^2} \,-\,1\Big)\,\,.
\eee

Substituting Eq. \eqref{A} into Eq.  \eqref{B}, we obtain that
\bee
\label{C}
\Bigg(8\pi G M T \,-\,\frac 4N \Bigg)\,e^{\delta M^2}\,=\,1\,-\,\frac 4N\,\,.
\eee

If we carry out the expansion of the exponential and, for simplicity, only the first two terms we write the general mass equation
\bee
\label{D}
M^3\,-\,\frac{1}{2\pi GTN}\,M^2\,+\,\frac 1\delta\,M\,-\, \frac{1}{8\pi GT\delta}\,=\,0 \,,
\eee

\ni and a possible solution for this system is 
\baa
\label{E}
M_1 &=& \frac 12 q\,N\,T \,, \noo
M_2 &=& \frac{2}{(4\pi GT)^2\,N\,(q-1)\Big(A\pm\sqrt{A^2\,-\,4B}\Big)} \,, \\
M_3 &=& \frac 12 \Big(A\pm\sqrt{A^2\,-\,4B}\Big) \,, \nonumber
\eaa

\ni where 
\baa
\label{F}
A &=& \frac 12 \Bigg(\frac{1-\pi G q N^2 T^2}{\pi G N T} \Bigg) \,,\noo
B &=& \frac{1}{(4\pi G T)^2 N (q-1)} \,,
\eaa

\ni and some conditions are quite obvious such as $A^2 > 4B$ and $A > \sqrt{A^2 - 4B}$ { since we cannot have imaginary and/or negative values for $M$}.  However, the limit $q \rightarrow 1$ introduced a divergence in $B$, which makes both $M_2$ and $M_3$ not physically viable.   Hence, we have that $M_1 =NT/2$ when $q \rightarrow 1$, which is the standard equipartition law.   Therefore, Eqs. \eqref{tbh}, \eqref{nis} and \eqref{eqis} represent the thermodynamical parts of the equipartition law in Eq.  \eqref{eqis} brought from IS.

The heat capacity, using Eqs. (\ref{isbh}) and (\ref{heat-capacity}), is
\begin{eqnarray}
\label{cis}
C_{IS}=- \, \frac{ \frac{8 \pi G M^2}{q} \, e^{4 \pi G M^2 \left( \frac{q-1}{q}\right)}}{1+8 \pi G M^2 \left(\frac{q-1}{q}\right)}  \,\,,
\end{eqnarray}

\ni  and when we make $q=1$ in Eq. (\ref{cis}) we recover the usual value of the Schwarzschild BH heat capacity written in Eq.  (\ref{h-capacity}).

From Eq.  (\ref{tbh}) we can derive a temperature variational condition \cite{ks} such that
\begin{eqnarray}
\label{dtext}
\frac{\partial T}{\partial M}\Bigg|_{M=M_{ext}} = 0 \; \qquad \Rightarrow \;\qquad 8 \pi G M^2_{ext} = \frac{q}{1-q}\,\,.
\end{eqnarray}

\ni So, substituting Eq.  (\ref{dtext}) into Eq. (\ref{tbh}) we can write the Hawking temperature as

\begin{eqnarray}
\label{hte}
T =  \frac{(1-q)\,  M_{ext}^2}{M}\,  \, \,\exp\Bigg( \frac{M^2}{2M^2_{ext}} \Bigg) \,.
\end{eqnarray}

\ni  Making use of Eq.  (\ref{dtext}) and only after that, using  $q=1$ into Eq. (\ref{hte}) (with $q$) then we recover the usual Hawking temperature, Eq.  (\ref{ts2}). In Fig. 3 the temperature, Eq.  \eqref{hte}, has been plotted as a function of mass $M$ for $q=1/2$, and from Eq.  (\ref{dtext})
and making $G=1$ we have $M_{ext}\cong 0.2$.

\begin{figure}[H]
	\centering
	\includegraphics[width=8.0cm]{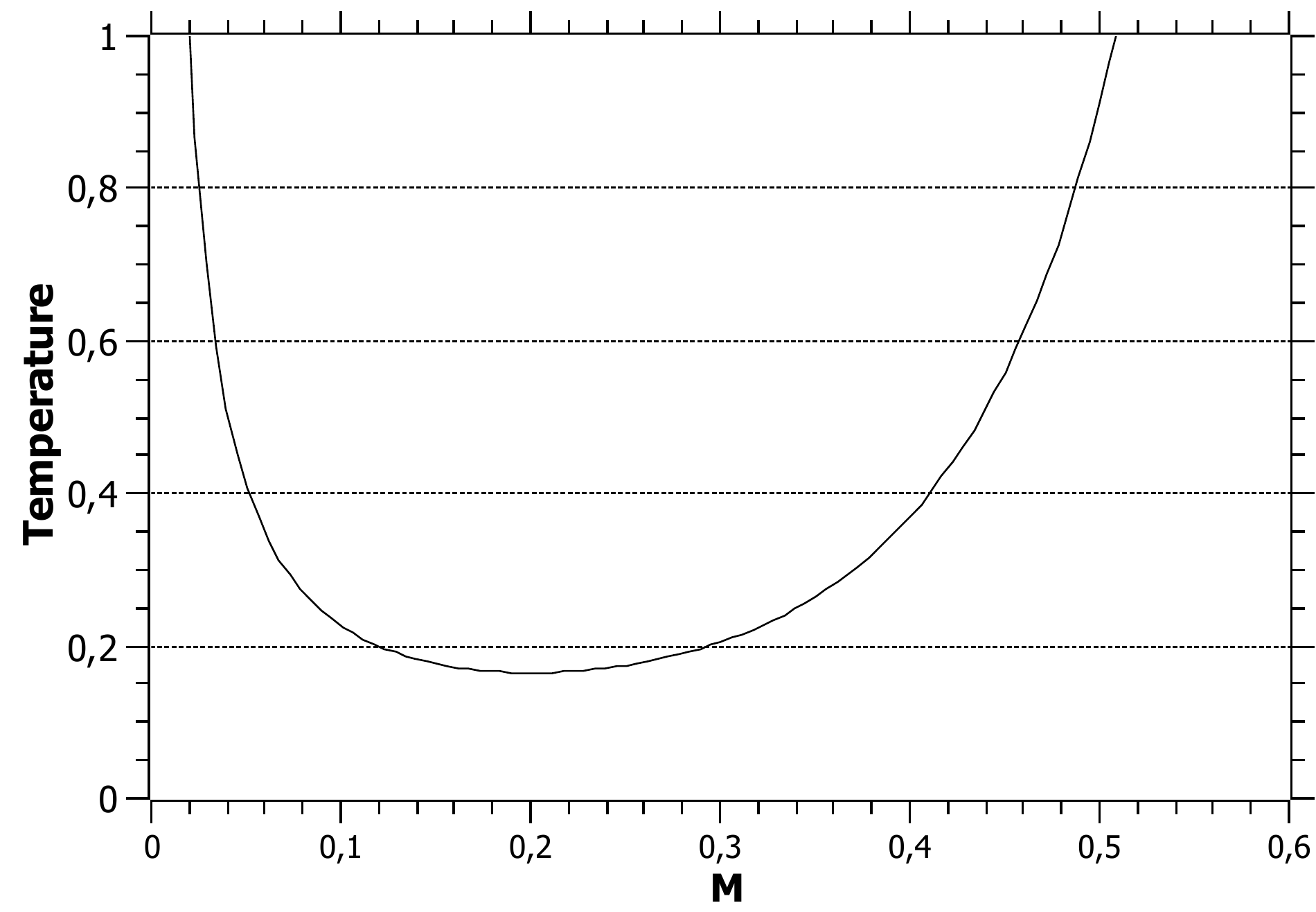}
	\caption{BH temperature of IS, Eq. (\ref{hte}), as a function of mass $M$ for $q=1/2$
		and $M_{ext}=0.2$.}
\end{figure}

\ni From Fig. 3 we can observe that for $M=0.2$ we have the minimum value of the temperature.

On the other hand, Eq.  (\ref{dtext}) together with the normalization condition, Eq.  (\ref{inn}), determines that the allowed values for the q-parameter, since the term $8 \pi G M^2_{ext}$ is always positive, is
\begin{eqnarray}
\label{ctext}
 0 < q < 1  \,.
\end{eqnarray}

\ni Due to Eq. (\ref{tbh}), the temperature is always positive, as it should be, if the condition (\ref{ctext}) is obeyed. So, using Eq.  (\ref{dtext}) in (\ref{cis}) we can write the heat capacity as
\begin{eqnarray}
\label{cvext}
C^{ext}_{IS} =  \frac{M^2 }{(q-1) \Big(M^2_{ext}-M^2\Big)}  \exp\Bigg({- \frac{M^2}{2 M^2_{ext}}}\Bigg) \,.                         
\end{eqnarray}

\ni From Eqs. (\ref{cvext}) and (\ref{ctext}) we can observe that for $ M < M_{ext}$ the heat capacity of system is negative. Consequently the BH has an unstable thermal equilibrium. For $ M > M_{ext}$ the heat capacity of system is positive and the BH is thermally stable. In Fig. 4, the heat capacity, Eq.  (\ref{cvext}), has been plotted as a function of mass $M$ for $q=1/2$ and $M_{ext}=0.2$. So, we can observe that for $M=M_{ext}=0.2$ the heat capacity diverges. This result can indicate a possible phase transition between the thermally unstable phase and the thermally stable phase of the BH in
the IS theory.

\begin{figure}[H]
	\centering
	\includegraphics[width=8.cm]{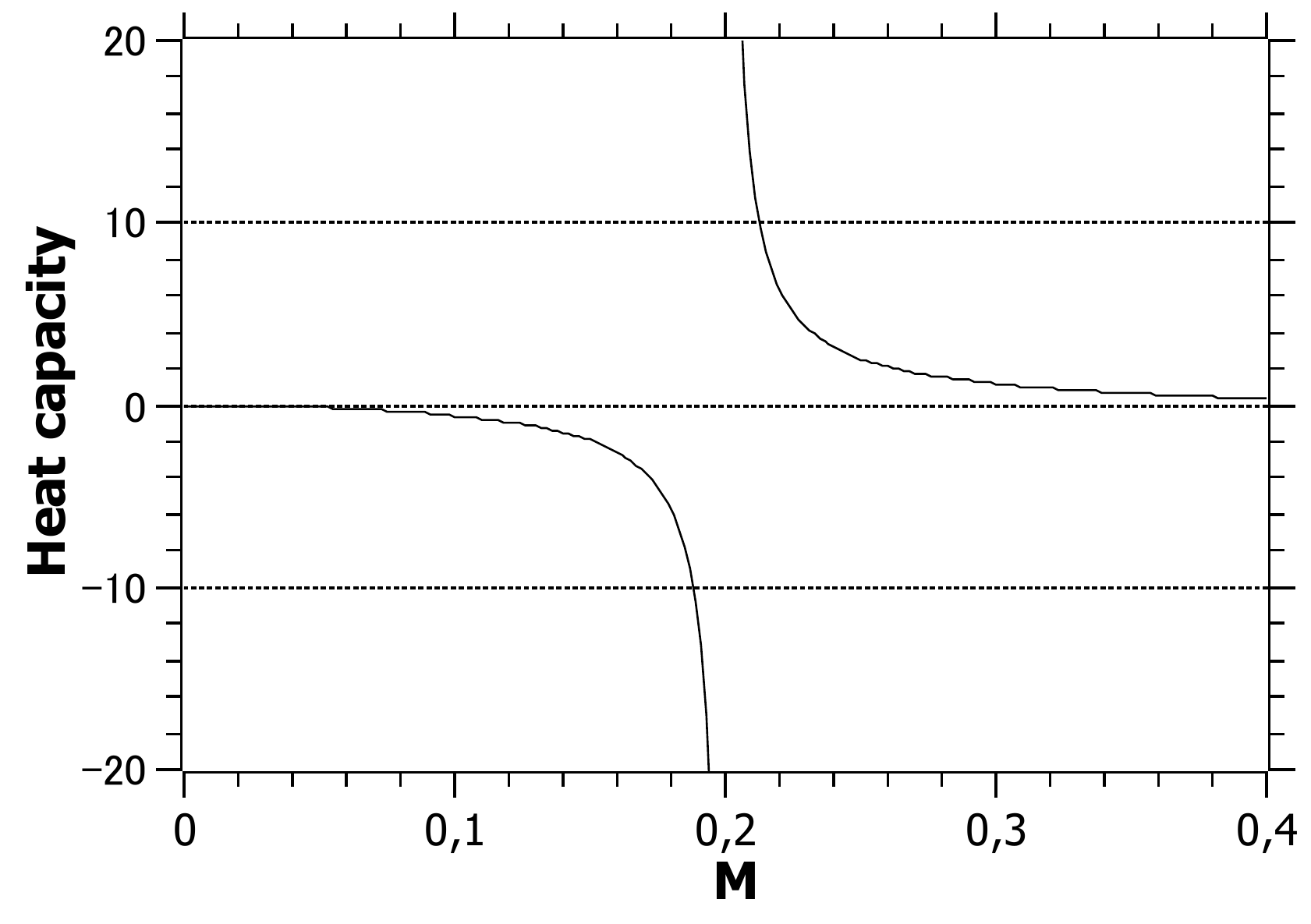}
	\caption{BH heat capacity of IS, Eq. (\ref{cvext}), as a function of mass $M$. We have used $q=1/2$ 
		and $M_{ext}=0.2$.}
\end{figure}

\section{Conclusions}

{ In this work we have investigated the effect of MRE and IS in the framework of BH thermodynamics.  
Using the variational conditions for the temperature function, Eqs.  (\ref{vctr}) and (\ref{dtext}), we could write the BH heat capacities in more suitable forms which are Eqs.  (\ref{crev}) and (\ref{cvext}). These equations exhibit divergent points when $M=M_{ext}$. These results mean that it is possible for phase transitions to occur, a fact that can not be noticed when we use BG theory. Consequently we can have  BHs thermally stable phases. Therefore, we can conclude that these interesting results point out the relevance of using non-Gaussian entropies, like both MRE and IS, in the analysis of BH thermodynamics.}

\section*{Acknowledgments}

\ni The authors are grateful to the anonymous referee for useful comments.   The authors also thank CNPq (Conselho Nacional de Desenvolvimento Cient\' ifico e Tecnol\'ogico), Brazilian scientific support federal agency, for partial financial support, Grants numbers  406894/2018-3 (Everton M. C. Abreu) and 303140/2017-8, 307153/2020-7 (Jorge Ananias Neto).


\begin{thebibliography}{99}

\bibitem{swh} S. W. Hawking, Commun. Math. Phys. 43 (1975) 199.

\bibitem{jdb} J. D. Bekenstein, Phys. Rev. D 7 (8) (1973) 2333.

\bibitem{tsallis} C. Tsallis, J. Stat. Phys. 52 (1988) 479.

\bibitem{levy} P. A. Alemany and D. H. Zanette, Phys. Rev. Lett. 75 (1995) 366.

\bibitem{turb} C. Anteneodo and C. Tsallis, J. Mol. Liq. 71 (1997) 255.

\bibitem{sys} C. Tsallis, Chaos, Soliton and Fractals 13 (2002) 371.

\bibitem{sa} R. Silva and J. S. Alcaniz, Physica A 341 (2004) 208.

\bibitem{eu} J. Ananias Neto, Physica A 391 (2012) 4320; E. M. C. Abreu, J. Ananias Neto, A. C. R. Mendes and Wilson Oliveira, Physica A 392 (2013) 5154.

\bibitem{maji} A. Majhi, Phys. Lett. B 775 (2017) 32.

\bibitem{mora2}  M.Tavayef, A. Sheykhi, Kazuharu Bamba and H. Moradpour, Phys. Lett. B 781 (2018) 195.

\bibitem{renyi}  A. Rényi, Probability Theory (North-Holland, Amsterdam, 1970).

\bibitem{ci} V. G. Czinner and H. Iguchi, Pys. Lett. B 752, 10 (2016) 306.

\bibitem{ko} N. Komatsu, Eur. Phys. J. C, 77 (2017) 229.

\bibitem{many} H. Moradpour, A. Bonilla, E.M.C. Abreu and J.Ananias. Neto, Phys. Rev. D
96 (2017) 123504; H. Moradpour, A. Sheykhi, C. Corda and I.G. Salako, Phys. Lett. B
783 (2018) 82;  E.M.C. Abreu, J.Ananias. Neto, Ed\'esio M. Barboza, Jr., Albert C. R. Mendes and Br\'aulio B. Soares,
MPLA Vol. 35, No. 32 (2020) 2050266.

\bibitem{inco1} Q. A. Wang, Chaos, Soliton and Fractals 12 (2001) 1431.

\bibitem{lbs} J. A. S. Lima, J. R. Bezerra and R. Silva, Chaos, Solitons and Fractals 19 (2004) 1095.

\bibitem{todos} 
M. Pezeril, A. L. M\'ehaut\'e and Q. A. Wang, Physica A 340 (2004) 117;
Q. A. Wang, Eur. Phys. J. B 26 (2002) 357; Q. A. Wang, Eur. Phys. J. B 31 (2003) 75.


\bibitem{nos1}   E. M. C. Abreu, J. Ananias Neto, E. M. Barboza and B. B. Soares, EPL 127 (2019) 10006.

\bibitem{nos2}     E. M. C. Abreu, J. Ananias Neto, E. M. Barboza, A. C. R. Mendes and B. B. Soares, Mod. Phys. Lett. A, vol 35 (2020) 2050266-1.



\bibitem{nos3}    E. M. C. Abreu, J. Ananias Neto and E. M. Barboza, EPL 130 (2020) 4, 40005.

\bibitem{nos4}    E. M. C. Abreu and J. Ananias Neto, Eur. Phys. J. C, 80 8 (2020) 776.

\bibitem{nos5}     E. M. C. Abreu and J. Ananias Neto, Phys. Lett. B 807 (2020) 135602.


\bibitem{ks}    K. Mejrhit and S.-E. Ennadifi, Phys. Lett. B 794 (2019) 45.

\bibitem{mora}    H. Moradpour, A. H. Ziaie and M. Kord Zangeneh, Eur. Phys. J. C, 80 8 (2020) 732.


\end{thebibliography}
\end{document}